\documentclass[showpacs,preprintnumbers]{revtex4}
\usepackage{graphicx}
\usepackage{dcolumn}
\usepackage{bm}

\def\bea{\begin{eqnarray}}
\def\eea{\end{eqnarray}}
\def\sst{\scriptscriptstyle}
\begin{document}
\title{A causal model for a closed universe}
\author{Mauricio Cataldo}
\altaffiliation{mcataldo@ubiobio.cl} \affiliation{Departamento de
F\'\i sica, Facultad de Ciencias, Universidad del B\'\i o-B\'\i o,
Avenida Collao 1202, Casilla 5-C, Concepci\'on, Chile.}
\author{Sergio del Campo}
\altaffiliation{sdelcamp@ucv.cl} \affiliation{Instituto de F\'\i
sica, Facultad de Ciencias B\'asicas y Matem\'aticas, Pontificia
Universidad Cat\'olica de Valpara\'\i so, Avenida Brasil 2950,
Valpara\'\i so, Chile.}
\author{Paul Minning}
\altaffiliation{pminning@udec.cl} \affiliation{Departamento de
F\'\i sica, Facultad de Ciencias F\'\i sicas y Matem\'aticas,
Universidad de Concepci\'on, Casilla 160-C, Concepci\'on, Chile.}
\author{Francisco Pe\~na}
\altaffiliation{fcampos@ufro.cl} \affiliation{Departamento de
Ciencias F\'\i sicas, Facultad de Ingenier\'\i a, Ciencias y
Administraci\'on, Universidad de la Frontera, Avda. Francisco
Salazar 01145, Casilla 54-D, Temuco, Chile.\\}
\date{\today}
\begin{abstract}
We study a closed model of a universe filled with viscous fluid
and quintessence matter components. The dynamical equations imply
that the universe might look like an accelerated flat
Friedmann-Robertson-Walker (FRW) universe at low redshift. We
consider here dissipative processes which obey a causal
thermodynamics. Here, we account for the entropy production via
causal dissipative inflation.
\end{abstract}
\maketitle \preprint{GACG/5-2003}
\section{\label{sec:level1} Introduction}
Recent observational evidences suggest that the measured matter
density of baryonic and dark matter is significantly less than
one, i.e. its critical value. This implies that either the
universe is open or that there are some other matter components
which makes $\Omega_{total}\sim 1$. Combined measurements of CMB
temperature fluctuations and of the distribution of galaxies on
large scales strongly suggest the possibility of a flat
universe~\cite{Ostriker,Bernardis}, which is consistent with the
standard inflationary prediction\cite{Guth}. Recent measurements
of a Ia type distant supernova (SNe Ia )~\cite{Pe-etal,Ga-etal},
at redshift $z \sim 1$, indicate that the expansion of the present
universe is accelerated.

An initial interpretation was to consider that in the universe
there exists an important matter component that, in its most
simple description, has the characteristic of the cosmological
constant $\Lambda$, i.e. a vacuum energy density which contributes
to a large component of negative pressure, and thus accelerates
rather than decelerates the expansion of the universe. An
alternative interpretation is to consider quintessence (or dark
energy), in the form of a scalar field with a self-interacting
potential. In the literature various examples of such Q a
component have been considered starting from the pioneering
paper~\cite{CaDaSt}, which also considered the case of cosmic
strings. The non--minimally coupled scalar fields were considered
in Ref.~\cite{Faraoni} and the scalar--tensor theories in
Ref.~\cite{Bartolo}. The current status about dark energy can be
found in Ref.~\cite{Turner}.

On the other hand, in recent years important attention has been
received by cosmological models which consider bulk viscosity. For
instance, it was shown that the introduction of this kind of
viscosity into cosmological models can avoid the big bang
singularities~\cite{Murphy,Belinkskii,Golda}, and any contribution
from particle production may be modeled as an effective bulk
viscosity~\cite{Maartens}. The bulk viscosity typically arises in
mixtures of different species (as in a radiative fluid) or of the
same (but with different energies) fluids. The dissipation due to
bulk viscosity converts kinetic energy of the particles into heat,
and thus one expects it to reduce the effective pressure in an
expanding fluid. This fact may play a crucial role in the
inflationary era of the universe since it is interesting to know
whether dissipative effects could be strong enough to make a large
negative effective pressure leading to inflation.

Many, probably most, of the inflationary cosmological models
considered are of a flat FRW type, since they are spatially
isotropic and homogeneous. This implies that velocity gradients
causing shear viscosity, and temperature gradients leading to heat
transport, are absent. Thus any dissipation in an exact FRW
universe is scalar, and therefore may be modeled as a bulk
viscosity within a thermodynamical approach. These scalar
dissipative processes may be treated in cosmology via the theory
of general relativity, considering the bulk viscosity~\cite{4,5},
which is compatible with the homogeneity and isotropy assumptions
for the universe. These dissipative processes may play an
important role in the early universe, especially before
nucleosynthesis~\cite{Gron1}. Then we shall use the causal
thermodynamical theory for processes not in
equilibrium~\cite{Israel}. The stable and causal thermodynamics of
Israel and Stewart replaces satisfactorily the unstable and
non-causal theory of Eckart~\cite{Eckart} and Landau and
Lifshitz~\cite{Landau}.

It is well know that the inflationary universe model is necessary
for solving most of the cosmological puzzles, in the way that it
explains how a large class of initial states can evolve into a
unique final state that is consistent with our observed universe.
In most of inflationary FRW cosmological models not only the scale
factor $a(t)$ presents a positive acceleration $\ddot{a}>0$, but
also it is assumed from the very beginning that the geometry of
the universe is completely flat.

In the light of this approach an interesting question to ask is
whether this flatness may be due to a sort of compensation among
different components that enter into the dynamical equations. In
this respect, one of our goals is to address this question in a
frame of a simple model. In the literature we find some
descriptions along these lines. For instance, a closed model has
been studied with an important matter component whose equation of
state is given by $p = - \rho / 3$. There, the universe expands at
a constant speed~\cite{Ko}. Other authors, while using the same
astronomically observed properties for the universe, have added a
nonrelativistic matter density in which the total matter density
$\Omega_0$ is less than one, thus describing an open
universe~\cite{KaTo}. Also, a flat decelerating universe model has
been simulated~\cite{CrdCHe}. The common fact to all of these
models is that, even though the starting geometry were other than
that corresponding to the critical geometry, i.e. a flat geometry;
all these scenarios are, at low redshift, indistinguishable from a
flat geometry.

In this paper we wish to consider a closed universe model composed
of two matter components: one related to a viscous matter and the
other to quintessence matter. The geometry, together with these
matter components, confabulates in such a special way that it
gives rise to a flat accelerating universe scenario. In this
regard, we use the standard anzats (see Eqs. (\ref{factor de
escala}), (\ref{suposiciones}) and (\ref{anzats}) below) which
introduce five parameters in our model, where there exists a range
of these parameters that gives an appropriate value for the
entropy of the universe. The value obtained for the entropy in our
model agrees with the expected value obtained from usual
inflationary re-heating. In this way, as we will see, our model
accounts for the generally accepted entropy production, via warm
inflation\cite{warm1}. In this kind of model the radiation is
continuously produced by the decay of the inflaton scalar field.
In this way this field should be coupled to ordinary matter and,
contrary to usual inflation, primordial density fluctuations are
originated much more from thermal fluctuations than from quantum
fluctuations of the inflaton\cite{warm2}. At the end, the
parameters of the resulting model could be estimated by using
astronomical data.

The outline of the present paper is as follows: In Sec. II we
briefly review the system of gravitational field equations and the
transport equation for bulk viscosity. In Sec. III and IV a power
law FRW and de Sitter FRW causal cosmologies are considered. In
Sec. V some conclusions are given.

\section{The system of gravitational field equations
and the transport equation for bulk viscosity} Now we shall obtain
the field equations for the considered cosmological viscous model.
The energy-momentum tensor of the fluid with a bulk viscosity is
given by
\begin{eqnarray}
\label{1} T_{\mu \nu }=(\rho_{M}+ p_{M}+\pi )v_{\mu }v_{\nu }+(p_{
M}+\pi)g_{\mu \nu },
\end{eqnarray}
where $\rho_{M}$ and $p_{M}$ are the thermodynamical density and
the pressure of the fluid, while $\pi$ is the bulk viscous
pressure, and $v^{\mu}=\delta_{0}^{\mu}$.

The contribution to the energy-momentum tensor related to the
classical scalar field $\phi $, which is minimally coupled to
gravity, becomes given by:
\begin{eqnarray} \label{2} T_{\mu \nu}=\phi_{,\mu}
\phi_{,\nu}-g_{\mu
\nu}\left(\frac{1}{2}\phi_{,\alpha}\phi^{,\alpha}+V(\phi)\right),
\end{eqnarray}
where $V(\phi )$ is the scalar potential.

Now we consider the FRW metric
\begin{eqnarray}
ds^2=-dt^2+a^2(t)\left(\frac{dr^2}{1-kr^{2}}+r^{
2}(d{\theta}^2+sin^{2}\theta \, d\phi^{2})\right).
\end{eqnarray}
>From Eq.~(\ref{1}) and~(\ref{2}) we have the gravitational field
equations:
\begin{eqnarray}
\label{00FRW} 3H^2 + 3\frac{k}{a^2}=\kappa \left(
\frac{1}{2}{\dot{\phi}}^2 + V + \rho_{\sst M}\right) \eea \bea
2\dot{H}+3H^2+\frac{k}{a^2}=\kappa
\left(-\frac{1}{2}{\dot{\phi}}^2 + V - p_{\sst M} - \pi \right)
\label{pi}
\end{eqnarray}
where $H=\frac{\dot{a}(t)}{a(t)}$ and $k$ is the curvature
parameter and $\kappa=8 \pi G$, with $G$ the Newtonian
gravitational constant.

The fluid conservation equation is given by
\begin{eqnarray}
\label{(6)} \dot{\rho}_{\sst M}+3H(\rho_{\sst M}+p_{\sst M}+\pi)=0
\end{eqnarray}
and the equation for the scalar field by
\begin{eqnarray}
\ddot{\phi}+3H\dot{\phi}+\frac{\partial V}{\partial \phi}=0.
\end{eqnarray}

The energy density associated with the scalar field is given by
\begin{eqnarray}
\label{rho phi} \rho_{_{\phi}}=\frac{1}{2}{\dot{\phi}}^{2}+V(\phi)
\end{eqnarray}
and the pressure by
\begin{eqnarray}
\label{p phi} p_{_{\phi}}=\frac{1}{2}{\dot{\phi}}^{2}-V(\phi).
\end{eqnarray}

The scalar field has a state equation defined by
$p_{_{\phi}}=w\rho_{_{\phi}}$ and satisfies the conservation
equation
\begin{eqnarray}
\label{conservacion del campo escalar}
\dot{\rho}_{_{\phi}}+3H(\rho_{_{\phi}}+p_{_{\phi}})=0.
\end{eqnarray}

>From Eq.~(\ref{00FRW}) and the weak energy condition
$\rho_{_{\phi}}\geq 0$ we conclude that, for desired flatness
simulation, i.e. Eq.~(\ref{00FRW}) becoming the well known
Friedmann Equation for a flat FRW universe
\begin{eqnarray}
\label{00 simulada} 3H^2=\kappa \rho_{_{M}},
\end{eqnarray}
we must take $k=1$ in the field equations. Thus
\begin{eqnarray}
\label{rho phi de a} \kappa \rho_{_{\phi}} = 3/a^2.
\end{eqnarray}
>From here we have that $\kappa
\dot{\rho}_{_{\phi}}=-3\dot{a}/a^3=-6H/a^2$ and, substituting into
Eq.~(\ref{conservacion del campo escalar}), obtain that
$w=-\frac{1}{3}$. Notice that this result is independent of the
explicit expression for the scale factor as a function of $t$.

At this point we should notice that our model is intrinsically
closed, therefore we should stress that our model is completely
different with the flat standard Einstein model which
characterizes with $\Omega_M = 1$ and is ruled out by the currents
astronomical observations. In fact, the curvature term (which is
intrinsically geometric) cannot be eliminated by a simple
redefinition of the scalar field $\phi$, since the curvature is a
geometrical property, which follows directly from the metric
tensor and which enters into the FRW line elements. These two
models become indistinguishable only for low redshifts and are
similar to the cases studied in Refs.~\cite{KaTo, CrdCHe, Varios}.

Another point that we would like to stress here is that almost all
physical quantities are obtained in terms of the scale factor and
its derivatives. From Eq.~\ref{rho phi de a} we get
$\rho_{_{\phi}}$ as a function of $a$. Since $w = - 1/3$, then
$\dot{\phi}^2 = V(\phi) = 2/\kappa a^2$. Also, from Eq.~\ref{pi}
we get that $\kappa \pi = -23 \dot{H}^2 -3 \gamma H^2$, where we
have used the state of equation $p_M = (\gamma -1)\rho_{_M}$ for
the matter component and Eq.~\ref{00 simulada}. As we will see and
in order to obtain an explicit expression for the scalar potential
$V$ as a function of the scalar field $\phi$, we need the temporal
dependence of the scale factor. In this respect we will take a
power law and an exponential inflationary solutions.

>From now on we shall use the simplest case of scalar dissipation
due to bulk viscosity $\xi$, modeling cosmological dissipation
from the consideration of nonequilibrium thermodynamics via
Israel-Stewart theory. The bulk viscous pressure $\pi$ is given by
the transport equation (linear in $\pi$):
\begin{eqnarray}
\label{8} \tau \dot{\pi}+\pi =-3 H \xi -\frac{1}{2} \tau \pi
\left(3H+\frac{\dot{\tau}}{\tau}-\frac{\dot{\xi}}{\xi}-\frac{\dot{T}}{T}\right),
\end{eqnarray}
where $\tau$ is the relaxation time (which removes the problem of
infinite propagation speeds), $\xi$ the coefficient of bulk
viscosity and $T$ is the temperature of the fluid. In the
non-causal formulation $\tau=0$ and then Eq.~(\ref{8}) has a
simple form $\pi=-3 H \xi$.

Following~\cite{Belinskii,4} we shall take the different
thermodynamic quantities to be simple power functions of the
density $\rho_{_{M}}$,
\begin{eqnarray}
\label{suposiciones} \xi = \alpha {\rho_{\sst M}}^{m}, \;\;\; T=
\mu {\rho_{\sst M}}^r \;\;\; and \;\;\; \tau
=\frac{\xi}{\rho_{_{M}}}= \alpha {\rho_{\sst M}}^{m-1}
\end{eqnarray}
with $\alpha$,$\mu$,$m$ and $r$ positive constants. The expression
for $\tau$ is used as a simple procedure to ensure that the speed
of viscous pulses does not exceed the speed of light. For an
expanding cosmological model the constant $m$ should be positive
and
\begin{eqnarray}
\label{condicion para tau} \tau > H^{-1}
\end{eqnarray}
in order to have a physical behaviour for the coefficient of
viscosity $\xi$ and the relaxation time $\tau$ ~\cite{4,Banerjee}.

\section{Power law FRW causal inflationary cosmology}
Now we are interested in power law inflationary models; thus we
shall take the scale factor in the form
\begin{eqnarray}
\label{factor de escala} a(t)=a_0\left(\frac{t}{t_0}\right)^{n},
\end{eqnarray}
where $a_0$ and $n$ are constant parameters, with $n
>1$. Then,
using Eq.~(\ref{factor de escala}) we obtain
\begin{eqnarray}
\label{constante de Hubble} H=\frac{n}{t},
\end{eqnarray}
\begin{eqnarray}
\kappa \rho_{_{\phi}}=
\frac{3}{{a_{0}}^2}\left(\frac{t_0}{t}\right)^{2n}
\end{eqnarray}
and
\begin{eqnarray}
\kappa p_{_{\phi}}=
-\frac{1}{{a_{0}}^{2}}\left(\frac{t_0}{t}\right)^{2n}
\end{eqnarray}
for the Hubble constant, mass density and pressure of the field
$\phi$ respectively.

In this case the solution for the scalar field and its potential
are given by
\begin{eqnarray}
\phi(t)=\phi_{0}\left(\frac{t_0}{t}\right)^{n-1}
\end{eqnarray}
and \begin{eqnarray}
V(\phi)=V_{0}\left(\frac{\phi}{\phi_0}\right)^{\frac{2n}{n-1}},
\end{eqnarray}
respectively, where $\phi_0=\phi(t_0)=\pm
\frac{\sqrt{2/\kappa}}{1-n}\frac{t_0}{a_0}$ and
$V_{0}=\frac{2}{\kappa{a_{0}}^2}$.

We shall consider for the fluid a barotropic equation of state
\begin{eqnarray}
\label{stateeq} p_{_{M}}=(\gamma-1)\rho_{_{M}},
\end{eqnarray}
where $0 \leq \gamma \leq2$.

Then for the power law expansion~(\ref{factor de escala}) and from
eqs.~(\ref{(6)}), (\ref{8}) and~(\ref{suposiciones}) (with the
standard thermodynamic relation for the temperature of a
barotropic fluid $r=(\gamma-1)/\gamma$) we obtain
\begin{eqnarray} \label{27}
\ddot{\rho}_{\sst M}+(3n+1)\frac{\dot{\rho}_{\sst
M}}{t}+\frac{1}{\alpha}{\rho}^{1-m}_{\sst M}\dot{\rho}_{\sst
M}+\frac{3n \gamma}{\alpha t}{\rho}^{2-m}_{\sst M} \nonumber
\\
-9n^2\left(1-\frac{\gamma}{2}\right)\frac{\rho_{\sst
M}}{t^2}-\frac{(2\gamma-1)}{2\gamma}\frac{\dot{\rho}^{2}_{\sst
M}}{\rho_{\sst M}}=0.
\end{eqnarray}
Note that this equation may also be written in terms of $H$ using
Eq.~(\ref{00 simulada}).

{\bf Case $m\neq 1$:} For solving Eq.~(\ref{27}) we shall take the
following anzats~\cite{Banerjee}:
\begin{eqnarray}
\label{anzats} \rho_{_{M}}=\rho^{0}_{_{M}}
\left(\frac{t}{t_0}\right)^{(\frac{-1}{1-m})}.
\end{eqnarray}
For physical sense we must have $m<1$. Inserting~(\ref{anzats})
into Eq.~(\ref{27}) we obtain, for $m \neq 1$, that the constant
$\rho^{0}_{_{M}}$ is given by
\begin{eqnarray}\label{arbitrario}
\rho^0_{_M}=\left( \frac{\alpha}{t_0 (1-m)} \right) \times
\,\,\,\,\,\,\,\,\,\,\,\,\,\,\,\,\,\,\,\,\,\,\,\,\,\,\,\,\,\,\,\,
{}^{}
\\ \nonumber \left(
\frac{(2\gamma)^{-1}-3n(1-m)-9n^2(1-\gamma/2)(1-m)^2}{(1-3n\gamma(1-m
))}\right)^{1/(1-m)}.
\end{eqnarray}
Later on we shall assume that the parameters $\gamma$, $n$, $m$
and $\rho^0_{_{M}}$ are given; then we can find the parameter
$\alpha$. The pressure $p$ is obtained by using the equation of
state~(\ref{stateeq}).

Now from Eq.~(\ref{stateeq}) and Eq.~(\ref{(6)}) we obtain for the
bulk viscous pressure
\begin{eqnarray}\label{shear viscosidad}
\pi= -\frac{\dot{\rho}_{_{M}}}{3H}- \gamma \rho_{_{M}}
\end{eqnarray}
and substituting~(\ref{anzats}) and $p_{_{M}}$ into~(\ref{shear
viscosidad}) we obtain
\begin{eqnarray}
\label{final de viscosidad} \pi= -\left(\gamma-\frac{1}{3n(1-m)}
\right) \rho^0_{_{M}} \left(\frac{t}{t_0}\right)^{-1/(1-m)}.
\end{eqnarray}
>From here we see that $\pi<0$ if $\gamma>1/(3n (1-m))$, $\pi > 0$
if $\gamma<1/(3n (1-m))$ and $p=0$ if $m=1-1/(3n\gamma)$.

The entropy and the temperature for local equilibrium satisfy the
Gibbs equation
\begin{eqnarray} \label{32}
TdS=(\rho_{M}+p_{M})d\left(\frac{1}{N}\right)+\frac{1}{N}
d\rho_{M}
\end{eqnarray}
where $N$ is the number density, and satisfy the conservation
equation
\begin{eqnarray} \dot{N}+3HN=0
\end{eqnarray}
from which we get the solution
\begin{eqnarray} \label{34}
N(t)=N_{0}\left(\frac{t}{t_{0}}\right)^{-3n}
\end{eqnarray}
or equivalently $N a^{^{3}}(t)= const$.

Using~(\ref{32}) and~(\ref{34}) we obtain the well known evolution
equation for the entropy (neglecting the heat flux and the shear
viscosity):
\begin{eqnarray} \label{35}
\dot{S}=-\frac{3H\pi}{NT}.
\end{eqnarray}
>From eqs.~(\ref{constante de Hubble}), (\ref{suposiciones}),
(\ref{anzats}), (\ref{final de viscosidad}) and~(\ref{34}) we have
\begin{eqnarray} \label{Spunto}
\dot{S}=\frac{3n\gamma-1/(1-m)}{N_{_{0}}\mu t_{_{0}}}
(\rho^0_{_{M}})^{1/\gamma}
\left(\frac{t}{t_0}\right)^{(3n-1)-1/\gamma(1-m)}.
\end{eqnarray}
For a reasonable physical behaviour we must satisfy $\dot{S}>0$,
which implies the condition $m<1-1/\gamma(3n-1)$.

The total entropy $\Sigma$ in a comoving volume is defined by
$\Sigma=S N a^{^{3}}(t)$. Then, by considering Eqs.~(\ref{34})
and~(\ref{35}) we may write the growth of total nondimensional
comoving entropy over a proper time interval from $t_i$ until
$t_f$ as~\cite{4}:
\begin{eqnarray}
\label{total comoving entropy}
\Sigma_f-\Sigma_i=-\frac{3}{k_{_{B}}} \int^{t_f}_{t_i} \frac{\pi H
a^{^{3}}(t)}{T} dt.
\end{eqnarray}

Now taking respectively $t_i$ and $t_f$ as the beginning and the
exit time for the inflation period of the universe, and from
eqs.~(\ref{factor de escala}), (\ref{constante de Hubble}),
(\ref{suposiciones}), (\ref{anzats}), (\ref{final de viscosidad})
and~(\ref{total comoving entropy}), we obtain for the increase in
total nondimensional entropy in the comoving volume $a^{^{3}}(t)$
the following expression:
\begin{eqnarray}
\label{Dedo} \Sigma_f-\Sigma_i=\frac{\gamma a^3_0 \,
(\rho^0_{{_M}})^{1/\gamma}}{k_{_{B}} \mu} \times \nonumber \\
\left[\left(\frac{t_f}{t_0} \right)^{3n-\frac{1}{\gamma(1-m)}} -
\left(\frac{t_i}{t_0} \right)^{3n-\frac{1}{\gamma(1-m)}} \right].
\end{eqnarray}
The typical values for the beginning and ending times of inflation
are $t_i \approx 10^{-35}$ s and $t_f \approx 10^{-32}$ s
respectively. In our following numerical considerations we shall
take the reference time $t_0$ equal to the ending one, i.e.
$t_0=t_f=10^{-32}$ s. Considering that the universe at the end of
the inflation period exits to the radiation era, we can constrain
some of the constants of integration of the above formulae.
Effectively, we know that the temperature of the universe at the
beginning of the radiation era was $T=10^{14} \, GeV = 1.16 \times
10^{27}$ K~\cite{Blau}. Then we have that the temperature of the
end of inflation must be $T_f=1.16 \times 10^{27}$ K. On the other
hand we know that during this radiation period $\rho=a_r\, T^4$ is
valid, where $a_r=\frac{\pi^2k^4_{_B}}{15 c^3 \hbar^3}=7.56 \times
10^{-15} \, J \, m^{-3} K^{-4}$. From this we conclude that at the
end of the inflation period (or at the beginning of the radiation
era) we have $\rho_{_{f}} \approx 10^{93} \, J \, m^{-3}$. This
implies that in Eq.~(\ref{suposiciones}) $r=1/4$ for the
temperature, i.e. $\gamma=4/3$.

Then we can summarize these typical values for the inflation
period as~\cite{4,Kolb,Cheng}:
\begin{eqnarray}
\label{values} t_i \approx 10^{-35} \, s; \,\,\, t_f \approx
10^{-32} \, s; \,\,\, \,\,\, a_i \approx c t_i, \nonumber \\ T_f
\approx 10^{27} \, K, \,\,\, \gamma=4/3, \,\,\, \rho \approx
\times 10^{93} J/m^3.
\end{eqnarray}
The e-folding parameter $Z=ln[a(t_f)/a(t_i)]$ for the power law
inflation~(\ref{factor de escala}) takes the form
\begin{eqnarray}
\label{Z para power law} Z=n \, ln\left(\frac{t_f}{t_i} \right).
\end{eqnarray}
It is well known that for solving the problems of the standard
model in cosmology we must have $Z \approx 60-70$. Thus
from~(\ref{suposiciones}), (\ref{Dedo}), (\ref{values})
and~(\ref{Z para power law}) we have
\begin{eqnarray}
\label{Sigma para tn} \Sigma_f-\Sigma_i \approx \frac{4 a_i \,
e^{3Z} (\rho_{_{M}}^f)^{3/4}}{3 k_{_{B}} \mu} \times 10^{26}
\left(1- (10^{-3})^{3n-\frac{3}{4(1-m)}} \right),
\end{eqnarray}
where we have considered the relation $a^3_f=a_i^3 e^{3Z}$
following from the definition of $Z$. We see from Eq.~(\ref{Sigma
para tn}) that, if the inequality $3n \gamma (1-m)>>1$ (as we have
obtained above) is satisfied (in this case $\Sigma_f
>> \Sigma_i$), we obtain the accepted value for the total entropy
in the observable universe~\cite{Blau,Kolb}
\begin{eqnarray}
\label{valor sigma} \Sigma \approx 10^{88}.
\end{eqnarray}
Thus this model can account for the generally accepted entropy
production~(\ref{valor sigma}) via causal dissipative inflation,
without re-heating.

Thus, we describe a dissipative accelerating period for the
universe which resembles a warm inflationary scenario. To see this
more clearly let us consider eq.(\ref{(6)}). Introducing the
scalar inflaton field $\chi$ in such a way that we define $\rho_M
\equiv \frac{1}{2}\dot{\chi}^2+V(\chi)$ and $p_M \equiv
\frac{1}{2}\dot{\chi}^2-V(\chi)$, where $V(\chi)$ is the effective
potential associated with the inflaton field, we obtain the
following expression:
\begin{eqnarray} \label{36} (3 H +
\Gamma)\dot{\chi} + V'(\chi) = 0,
\end{eqnarray}
where we have introduced $\Gamma$ such that $\Gamma \dot{\chi} =
3H \pi/\dot{\chi} $. This term describes the density energy
dissipated by the $\chi$ field into the thermalized bath. We have
also used the slow-roll-over condition given by $\ddot{\chi} \ll
(3 H + \Gamma)\dot{\chi}$ (or equivalently, $\dot{\rho}_{\sst
M}\ll +3H(\gamma \rho_{\sst M} +\pi)/(\gamma - 1)$). Expression
(\ref{36}) is the basic equation which describes a warm
inflationary universe model~\cite{warm1}. One of the interesting
characteristics of this model is does not need an intermediate
reheating period as in usual inflationary models, since the
transformation of vacuum energy into radiation energy occurs
throughout the inflationary period and thus the inflationary epoch
smoothly terminates into a radiation dominated regime\cite{BrYa}.

At the end of the inflationary period, when the universe enters
into the radiation dominated period, we have found that the growth
of entropy which results is $\Sigma_f \approx 10^{88}$. This
result gives an answer to the total entropy problem, whose
question is why the total entropy in the observable part of the
universe is so large. Certainly this problem is one of the various
``puzzles" found in the standard Big-Bang cosmological model.

A special case is obtained here when the parameter $m=1/2$. In
this case we obtain a constraint for each and all of the constants
except $\rho^0_{_{M}}$. Effectively, substituting the expression
$\rho_{_{M}}=\rho^0_{_{M}}(t_0/t)^{2}$ into Eq.~(\ref{27}) we
obtain
\begin{eqnarray}\label{form}
\alpha=\frac{\sqrt{3/\kappa}\, n
(3n\gamma-2)}{9(1-\gamma/2)n^2+6n-2/\gamma}.
\end{eqnarray}
Now the relaxation time should satisfy the
condition~(\ref{condicion para tau}). But $\tau=\alpha
\rho^{^{1/2}}_{_{M}}$ and considering~(\ref{00 simulada}) we get
$\alpha << \sqrt{3/\kappa}$. This finally implies that
\begin{eqnarray}
n >>1/\gamma,
\end{eqnarray}
which is satisfied for a power law inflationary universe, since
there $n$ is greater than one. We could get an estimation of the
parameter $n$ by using the result~(\ref{valor sigma}) together
with Eq.~(\ref{Sigma para tn}) in which $m=1/2$. We get that $n
\gtrsim 1.6$ for $Z \lesssim 70$.

{\bf Case $m=1$:} For this value of $m$ we have to start from
Eq.~(\ref{27}) which gives the solution~\cite{Banerjee}
\begin{eqnarray}
\rho_{_{M}}=\rho^0_{_{M}} e^{-2\gamma t/\alpha} \left(
\frac{t}{t_0}\right)^{-\gamma(3n+2)},
\end{eqnarray}
where $n^2=2\gamma/9$ and $\rho^0_{_{M}}$ is an arbitrary
constant. Unfortunately this solution implies a decreasing entropy
\begin{eqnarray}
S(t)=\frac{\gamma}{\mu_0 N_0} \left(\rho^{^{0}}_{_{M}}
\right)^{1/\gamma} \left( \frac{t_0}{t} \right)^2 e^{-2 t/\alpha}+
const
\end{eqnarray}
which is a non-physical behaviour.

\section{de Sitter FRW causal inflationary model}
We also can consider the de Sitter kind of inflation models, for a
closed FRW with flat dynamics. In this case the scale factor is
given by $a(t)=e^{H_0 t}$, where $H_0=const$, and the solution for
the scalar field and its potential is obtained from Eqs.(\ref{rho
phi}), (\ref{p phi}) and~(\ref{rho phi de a}) and is given by
\begin{eqnarray}
\phi(t)=\phi_0 \, e^{-H_0t}
\end{eqnarray}
and
\begin{eqnarray}
V(\phi)= V_0 \left(\frac{\phi}{\phi_0} \right)^{2},
\end{eqnarray}
respectively, where $\phi_0=\pm \sqrt{2/\kappa}\, / H_0$ and
$V_0=2/\kappa$.

Now we have $H=H_0$ or equivalently, by Eq.~(\ref{00 simulada}),
$\rho_{_{M}}=const$. Thus from~(\ref{(6)}) we obtain that the
viscous pressure is given by $\pi=-\gamma \, \rho_{_{M}}$ and then
from~(\ref{35}) we see that always $\dot{S} \geq 0$. Now
Eq.~(\ref{8}) yields $H=\gamma \, \rho^{^{1-m}}_{_{M}}/(3
\alpha(1-\gamma/2 ))$ and considering Eq.~(\ref{00 simulada}), we
have for $m\neq 1/2$:
\begin{eqnarray}
H=H_0=: \left[\frac{3^{^{m}} \, \alpha(2 -\gamma)}{2
\kappa^{^{m-1}} \gamma } \right]^{1/(1-2m)},
\end{eqnarray}
with
\begin{eqnarray}
\xi=\xi_0=\alpha \left(\frac{3}{\kappa}\right)^{^m} H^{^{2m}}_0,
\,\, T=T_0=\mu \left(\frac{3}{\kappa} H^2_0 \right)^{(\gamma-1)/
\gamma}.
\end{eqnarray}
For $m=1/2$ we conclude that $H=\tilde{H}$, where $\tilde{H}$ is
an arbitrary constant and
\begin{eqnarray}
\alpha=\frac{2 \gamma}{\sqrt{3 \kappa} (2-\gamma)}.
\end{eqnarray}
Here
\begin{eqnarray}
\xi=\alpha \sqrt{\frac{3}{\kappa}} \, \tilde{H}, \,\, T=T_0=\mu
\left(\frac{3}{\kappa} \tilde{H}^2_0 \right)^{(\gamma-1)/ \gamma}.
\end{eqnarray}
It is easy to show that in both cases, i.e. for any value of $m$,
we have that the relaxation time is given by
\begin{eqnarray}
\tau=\frac{2 \gamma} {3 H \,(2-\gamma)}.
\end{eqnarray}
Then from Eq.~(\ref{condicion para tau}) we have
\begin{eqnarray}
\gamma > \frac{6}{5}
\end{eqnarray}
in accordance with the results obtained in~\cite{4}.

Now considering that $a(t)=e^{H \, t}$, $Z=H(t_f-t_i)$ and
from~(\ref{total comoving entropy}) we obtain
\begin{eqnarray}
\label{DedoH} \Sigma_f-\Sigma_i=\frac{\gamma a^3_i \,
\rho^i_{{_M}}}{k_{_{B}} T} \left( e^{3Z}-1 \right),
\end{eqnarray}
where $c$ has been restored. From $Z \approx 60$
and~(\ref{values}) we have that $\Sigma_f >> \Sigma_i$ and obtain
that $\Sigma_f \approx 10^{88}$ in agreement with the accepted
value at the end of the period of inflation.

\section{Conclusions}
In this paper we have described a closed model universe composed
of a viscous matter and a dark energy components. The dark energy
together with the curvature term ``confabulate" so that they
result in the dynamical Friedmann Equation for a flat FRW
universe, as is expressed by Eq. (\ref{00 simulada}). On the other
hand, the viscous matter component gives rise to an inflationary
universe model at non-vanishing temperature.

We have used the anzats (\ref{factor de escala}),
(\ref{suposiciones}) and (\ref{anzats}) which introduce the
following five parameters: $n$, $\alpha$, $m$, $r$ and $\mu$.
There exists a range of these parameters that gives an appropriate
value for the entropy of the universe. In fact we saw that $n >
1$, $r =(\gamma-1)/\gamma$, where $\gamma$ is the equation of
state parameter, $m<1-1/\gamma(3n-1)$ for $m \neq 1$ and $\alpha$
being an arbitrary constant. For a given value of $\alpha$ we
would get a value of $\rho^{0}_{_{M}}$ by using
Eq.~(\ref{arbitrario}). From these ranges we have gotten an
appropriate value for the entropy of the universe $\Sigma \approx
10^{88}$. We could say the same for $m=1/2$, where now $\alpha$ is
given by Eq.~(\ref{form}), which should satisfy $\alpha <<
\sqrt{3/\kappa}$ and $n >> 1/\gamma$. Notice that, in this latter
case, the matter component corresponds to quintessence, since the
pressure becomes negative. Thus, the value obtained for the
entropy in our model agrees with the expected value obtained from
usual inflationary re-heating, which accounts for the generally
accepted entropy production, via warm inflation\cite{warm1}. The
radiation period, after inflation, is continuously produced by the
decay of the inflaton scalar field. In this way this field should
be coupled to ordinary matter and, contrary to usual inflation,
primordial density fluctuations are originated from thermal
fluctuations rather than from quantum fluctuations of the
inflaton\cite{warm2}.

Our results have been determined both for power-law (i.e. $a\sim
t^n$, $n>1$) and for de Sitter (i.e. $a\sim e^{H_0t}$, $H_0
=$const.) inflationary universe models.

This work was supported by CONICYT through grants FONDECYT N$^0$
1010485 (MC and SdC) and N$^0$ 1030469 (SdC and MC). Also it was
supported by Direcci\'on de Investigaci\'on de la Universidad del
B\'\i o-B\'\i o (MC), by grant 123.764-2003 of Direcci\'on de
Estudios Avanzados de la Pontificia Universidad Cat\'olica de
Valpara\'{\i}so (SdC), by grant 20228 of DIUFRO (FP) and by grant
98011023-1.0 of Direcci\'on de Investigaci\'on de la Universidad
de Concepci\'on (PM).

\end{document}